# Quantum chemical calculations to elucidate the electronic and elastic properties of topologically equivalent metal organic frameworks

Mo Xiong,[a] Neng Li,[*,a,b] George Neville Greaves,[a,b,c] Yuanzheng Yue,[a,d] and Xiujian Zhao[*,a]

[a]State Key Laboratory of Silicate Materials for Architectures, Wuhan University of Technology, Hubei Province, 430070, P. R. China

[b]Department of Materials Science &Metallurgy, University of Cambridge, Cambridge, CB3 0FS, UK.

[c]Institute of Mathematics, Physics and Computer Science, Department of Physics, Aberystwyth University, Aberystwyth SY23 3BZ, UK

[d]Section of Chemistry, Aalborg University, DK-9220 Aalborg, Denmark

* Corresponding authors' e-mails:

Neng Li, lineng@whut.edu.cn; Xiujian Zhao, opluse@whut.edu.cn

Phone number: +86-027-87652553; Fax number: +86-027-8788374

**ABSTRCT**: We have theoretically investigated the elastic properties of three topologically identical zeolitic imidazolate frameworks; ZIF-4, ZIF-62 and TIF-4, by means of ab initio calculations. The ZIFs are a subset of metal organic frameworks (MOFs), whose versatile functionality is providing exciting opportunities in solid state science. Understanding the relationship between structure and elastic properties of hybrid materials is making it possible to develop a new generation of functional materials. In this work we have determined the bulk $K$, shear $G$, and Young's moduli $E$, together with Poisson's ratio ($v$), and the universal elastic anisotropy index ($A^U$) from the computed elastic coefficients. Tensorial analysis of the elastic constants reveals unusual and highly anisotropic elastic behaviour. All three materials exhibit low Young's and shear moduli. In addition, their flexibility incorporates regions of negative Poisson's ratio (NPR) and negative linear compressibility (NLC). The elastic properties of these ZIF crystals are affected by the substituted organic linkages.

1. **Introduction**

Zeolitic imidazolate frameworks (ZIFs) are a unique class of porous hybrid materials known as a subset of metal-organic frameworks (MOFs) which have been studied extensively over the past decade due to their potential applications in gas sorption and separation, catalysis, drug delivery and sensing applications [1-4]. The structure of ZIFs consists of tetrahedral metal ions, typically $Zn^{2+}$ or $Co^{2+}$, linked by organic bridging ligands which are derived from imidazolate anions (Im = $C_3N_2H_3^-$) [5].

The mechanical properties of these porous frameworks are of vital importance in the optimization and fabrication of novel electrodes, thin-film sensors and microelectronic devices. Experimental work has revealed correlations between elastic properties and metal ion identity, organic linker, porosity and topology [6,7]. Computational work has verified these correlations, and in particular shown that several MOFs are unstable upon solvent removal, and that shear stresses lie behind the collapse of ZIFs with pressure [8,9], mirroring earlier work on the mechanical properties and structural stability of zeolites [10-13]. Yet to be fully understood however is the elastic moduli of mixed-linker zeolitic imidazolate frameworks, and how small changes in chemical composition may tune framework elastic response [6,14].

One ZIF of current interest is ZIF-4 [$Zn(Im)_2$], (Im=$C_3H_3N_2^-$) which crystallizes in the orthorhombic space group *Pbca* and possesses a *cag* topology like the mineral variscite $CaGa_2O_4$ [15-17]. The framework has recently been shown via recrystallization to lead to a "liquid-MOF", and melt at 865K [18,19]. Partial replacement of the Im ligand with bIm or mbIm (bIm = benzimidazole, $C_7H_5N_2^-$ and mbIm = methylbenzimidazole, $C_8H_7N_2^-$) is possible, resulting in ZIF-62 [$Zn(Im)_{1.75}(bIm)_{0.25}$] and TIF-4 [$Zn(Im)_{1.5}(mbIm)_{0.5}$] respectively. Their structures all include 8 nano-pores per cell connected by apertures of diameter about 2.1 Å. As shown in Fig.1(g)~(i), all contain 4 and 6 membered rings (4MR and 6MR).

The different ratios and identities of the organic ligands in the frameworks has a direct effect on the density and the porosity of the materials, although they have the same symmetry and topology. The porosity is described by the solvent accessible volume (SAV), which is a useful metric for applications in gas storage or guest encapsulation. Physical densities decrease in the order ZIF-4>ZIF-62>TIF-4, consistent with the size of the organic linker included (Table 1). The least dense material in the study, TIF-4, also possesses the smallest SAV, however, due to pore-blocking by the imidazole substituent.

In order to further investigate the role of the organic anions in influencing framework mechanical properties, we employed density functional theory (DFT) to study the detailed elastic behaviour of these ZIF-type materials. The full elastic constants tensor of these structures has been calculated. Based on these constants, the crucial elastic properties such as Young's modulus, shear modulus, Poisson's ratio, or linear compressibility have been determined.

## 2 Computational methodology

The first-principle method based, on density functional theory (DFT), was used in the geometry optimization and total energy calculations for the structures of ZIF-4, ZIF-62 and TIF-4, the projector augmented wave (PAW) method being implemented in the Vienna *ab initio* Simulation Package (VASP) code [20]. The electronic exchange and correlation potential were expressed using the Perdewe-Burkee-Ernzerh (PBE) version of the generalized gradient approximation (GGA) [21]. The cut-off energy for the plane wave basis set was 600 eV for all structures. Structural relaxations were performed to a tolerance of $1\times10^{-7}$ eV/atom in the total energy, yielding average forces of $1\times10^{-3}$ eV/Å. The stress level of the final equilibrium structure was less than 0.1 GPa. The relaxation of all the structures imposed no restrictions on the volume and lattice vectors. Since the unit cells were large containing nearly 400 hundred atoms, only one K-point at $\Gamma$(0, 0, 0) was used.

The single-crystal elastic constants, $C_{ij}$, of the elasticity matrix (tensor) were then computed by performing six finite distortions of the lattice and deriving the elastic constants from the strain-stress relationship with the code in VASP [22,23]. The mechanical properties were calculated via tensorial analysis implemented in the ELATE code [24]. The solvent accessible volumes (SAV) were calculated using the PLATON code [25].

## 3 Results and discussion

### 3.1 Crystal Structures

In order to compare with experimental densities of solvated crystals [6], the calculated density of desolvated crystals was augmented by the density of template molecules in the cavities, showing good agreement (Table 1).

### 3.2 Electronic properties

The calculated total density of states (TDOS) and partial density of states (PDOS) of three materials are shown in Fig. 2. A large band gap of ~2.5 eV indicates that they are insulators. From the TDOS, the occupied portion of the valence band (VB) consists of nearly all sharp peaks (unlike the conduction band which possesses few), consistent with their visible transparency. The PDOS provides insight into the atomic bonding. Above -6 eV, the Zn atoms interact strongly with N, C and H atoms. The VBs show a strong hybridization of N, C, H atoms, which populate the linkers of Im, bIm and mBIm for the three materials.

### 3.3 Single-crystal elastic stiffness coefficients $C_{ij}$

The elastic constants, $C_{ij}$, are obtained by solving the following equation:

$$\sigma_{ij} = \sum_{ij} C_{ij}\varepsilon_j$$

The mechanical properties, such as the Young's modulus ($E$), shear modulus ($G$), bulk modulus ($K$), Poisson's ration ($v$) and linear compressibility ($\beta$), can be calculated by tensorial analysis of the elastic constants $C_{ij}$. Due to being orthorhombic, these three materials all have nine independent elastic constants in the symmetric elastic matrix (stiffness tensor) with the following form [26,27]:

$$C_{orthorhombic} = \begin{pmatrix} C_{11} & C_{12} & C_{13} & 0 & 0 & 0 \\ & C_{22} & C_{23} & 0 & 0 & 0 \\ & & C_{33} & 0 & 0 & 0 \\ \vdots & & & C_{44} & 0 & 0 \\ & \ddots & & & C_{55} & 0 \\ symm. & & \cdots & & & C_{66} \end{pmatrix}$$

Among the diagonal coefficients, $C_{11}$, $C_{22}$ and $C_{33}$ indicate the stiffness along the three principal crystal axes, *a*, *b* and *c*, respectively, under a uniaxial strain. While $C_{44}$, $C_{55}$ and $C_{66}$ represent the stiffness against angular distortions when subjected to shear strains. The off-diagonal coefficients, $C_{12}$, $C_{13}$ and $C_{23}$, correspond to tensile-tensile coupling between any two orthonormal axes. The criterion for mechanical stability of orthorhombic crystal structures is given by [28]

$$C_{11} > 0, C_{22} > 0, C_{33} > 0, C_{44} > 0, C_{55} > 0, C_{66} > 0$$
$$[C_{11} + C_{22} + C_{33} + 2(C_{12} + C_{13} + C_{23})] > 0$$
$$(C_{11} + C_{22} - 2C_{12}) > 0, (C_{11} + C_{33} - 2C_{13}) > 0$$
$$(C_{22} + C_{33} - 2C_{23}) > 0$$

The compliance matrix, the inversion of the elasticity matrix (tensor), is used to calculate the mechanical properties of the three structures, in accordance with:

$$S_{ij} = C_{ij}^{-1}$$

All of the computed single-crystal elastic constants are summarized for each system in Table 2. Obviously, the elastic constants of all three structures satisfy the mechanical stability criteria, indicating that these structures are mechanically stable. The calculated principal elastic constants demonstrate that the chemical bonds along in the [010] direction is stronger than those in the other two directions ([100] and [001]) for these structures, and hence, $C_{22}$ values are larger than $C_{11}$ and $C_{33}$. These can be attributed to the fact that the organic ligands rings are aligned mainly in the [010] direction, as shown in the Fig.1.

### 3.4 Young's modulus (*E*)

Young's modulus (*E*) measures the stiffness of the structure under the unidirectional loading and is defined as the ratio of uniaxial stress over uniaxial strain along a unit vector **u**. We obtain $E(\mathbf{u})$ from the following tensorial equation:

$$E(\mathbf{u}) = \frac{1}{u_i u_j u_k u_l S_{ijkl}}$$

The 3D surfaces corresponding to a spherical plot of $E(\mathbf{u})$ and 2D plot for the xy, xz, and yz planes shown in Figs. 3(a)~(c) indicate that the Young's moduli of all three ZIF crystals are direction dependent. We found that ZIF-62 exhibits the greatest anisotropy of the three materials, with the corresponding $A_E$ ratio ($E_{max}/E_{min}$) of 7.08, i.e., two and three times larger than that of ZIF-4 and TIF-4 respectively.

The lowest $E$ value, ~1.07 GPa, 0.76 GPa and 2.73 GPa as shown in Table 3, for all three materials is found on the <100> axis, which corresponds to *a* axis. Maximum $E$ values for the frameworks were found to correlate well with the increasing linker size (Fig. 3 and Table 3).

Our theoretical results can be compared with experimentally determined Young's moduli ($E$) from templated single crystal nano-indentation. For example, the measured $E$ values for solvated ZIF-4, where $E\{111\}\approx\{100\}$ are reported to be ~4.6 GPa, and are larger than the calculated moduli for unsolvated $E\{111\}$ and $E\{100\}$ of 3.4 GPa and 1.07 GPa, respectively (Table 3). The discrepancy may be ascribed to the template adding stiffness, but also to inaccuracies inherent in the nano-indentation method.

**3.5 Shear modulus ($G$)**

The shear or rigidity modulus, $G$, is a measure of the stiffness of the structure under the effect of shear forces acting parallel to the material surface. It characterizes the resistance against shape change or angular geometrical distortions. If $G$ is large this implies a framework with a high structural 'rigidity' towards shear strain. It depends on an additional unit vector $\mathbf{v}$, which is perpendicular to vector $\mathbf{u}$, and is obtained from the following equation:

$$G(\mathbf{u},\mathbf{v}) = \frac{1}{u_i v_j u_k v_l S_{ijkl}}$$

The 3-D maximal and minimal shear representation surfaces are shown in Fig. 4. The shear modulus of ZIF-4 is relatively more anisotropic than those of ZIF-62 and TIF4, with the $G_{max}/G_{min}$ ratio ($A_G$) values 5.11, 4.54 and 2.29, respectively. All frameworks have very low minimum shear moduli, as shown in Table 3, and found experimentally in ZIF-8 [30]. The source can be understood by consideration of the 4-, 6-membered rings rectangular configuration in these structures, which are susceptible to shear forces as shown in Fig. 1. The tetrahedral environment of the Zn atoms and molecular linkages makes structure more compliant and unstable under shear forces, leading to the shear induced dynamic structural distortion [31].

### 3.6 Poisson's ratio (ν)

The Poisson's ratio is defined as the ratio of the transverse strain strain ($\varepsilon_t$) to the axial strain ($\varepsilon_a$) under the uniaxial deformation, which can be utilized to identify anomalous elastic behavior like auxeticity [10,11]. It can be obtained from the equations:

$$\upsilon(\mathbf{u},\mathbf{v}) = -\frac{u_i u_j v_k v_l S_{ijkl}}{u_i u_j u_k u_l S_{ijkl}}$$

As it is also a function of two unit vectors, the maxima and minima in 3-D surfaces are shown in Fig.5. A summary of the maximum and minimum *ν* values are listed in Table 3. We noticed that TIF-4 displayed the highest $\nu_{max}$, while ZIF-62 exhibits the lowest.

All of these materials exhibit auxeticity in certain crystallographic orientations with the negative values of $\nu_{min}$ -0.497, -0.547 and -0.049 for ZIF-4, ZIF-62 and TIF-4 respectively. The negative Poisson's ratio (NPR) means that a positive axial strain leads to an 'abnormal' transverse expansion. These unusual elastic phenomena occur when the axial strain (loading direction) acts along the $<\bar{1}10>$ axis. Our results show that the frameworks with *cag* topology may mostly exhibit the auxeticity, generally ascribed to directional re-entrant rings [10].

### 3.7 Linear compressibility (*β*)

The anisotropic compressibility *β*(**u**) represents the compression along an axis upon isostatic compression and can be obtained from the equation:

$$\beta(\mathbf{u}) = u_i u_j S_{ijkl}$$

The 3-D and 2-D anisotropic surfaces are shown in Fig.6, with the corresponding magnitudes summarized in Table 3. All the materials exhibit negative linear compressibility, along *c* axis for ZIF-4 and ZIF-62.

More interesting, the maximum linear compressibility of ZIF-62 is an order of magnitude higher than the other frameworks discussed, while the minimum value is two orders of magnitude higher. We notice that negative Poisson's ratio (NPR) and negative linear compressibility (NLC) simultaneously occur in these topologically equivalent frameworks, which is in agreement with the recent predictions for MIL-type materials [32].

**3.8 Averaged elastic properties**

The isotropic elastic properties for the texture-free polycrystalline ZIFs are discussed in this section. From the calculated elastic constants, we can obtain the isotropic averaged properties of the bulk (***K***), Young's (***E***) and shear (***G***) moduli via Voigt-Reuss-Hill (VRH) approximation [33]. For the orthorhombic system, Voigt's modulus can be written as:

$$K_V = \frac{1}{9}[C_{11} + C_{22} + C_{33} + 2(C_{12} + C_{13} + C_{23})]$$
$$K_R = (S_{11} + S_{22} + S_{33} + 2S_{12} + 2S_{13} + 2S_{23})^{-1}$$
$$G_V = \frac{1}{15}(C_{11} + C_{22} + C_{33} - C_{12} - C_{13} - C_{23}) + \frac{3}{15}(C_{44} + C_{55} + C_{66})$$
$$G_R = 15[4(S_{11} + S_{22} + S_{33}) + 3(S_{44} + S_{55} + S_{66}) - 4(S_{12} + S_{13} + S_{23})]^{-1}$$

The Hill bulk modulus (***B<sub>H</sub>***) and shear modulus (***G<sub>H</sub>***) can be given as:

$$K_H = \frac{1}{2}(K_R + K_V)$$
$$G_H = \frac{1}{2}(G_R + G_V)$$

Young's modulus ($E$) and Poisson's ratio ($v$) of the polycrystalline aggregate can be calculated via $K_H$ and $G_H$, as follows:

$$E_{VRH} = \frac{9K_H G_H}{3K_H + G_H}$$

$$v_{VRH} = \frac{3K_H - 2G_H}{2(3K_H + G_H)}$$

In order to comprehensively estimate the elastic anisotropy of these materials, the universal elastic anisotropy index $A^U$ is adopted, accounting for both the shear and bulk contributions [34]:

$$A^U = 5\frac{G^V}{G^R} + \frac{K^V}{K^R} - 6$$

A summary of the isotropic averaged properties of the bulk ($K$), Young's ($E$) and shear ($G$) moduli and the universal elastic anisotropy index $A^U$ are presented in Table 4. These values can be used to approximate the elastic properties of ZIF thin-film coatings and bulk extrudates, provided that their polycrystalline grain arrangements are in fact randomly oriented.

The bulk modulus is an isotropic measure that quantifies the resistance of the structure towards a volumetric strain ($\Delta V/V$) when subjected to a hydrostatic stress state (pressure). The values given in Table 4 are the Voigt-Reuss-Hill (VRH) averages. The Voigt values assume a uniform strain, the Reuss values correspond to a uniform stress, and the Voigt-Reuss-Hill values are the average of two. The bulk moduli increase with decreasing framework SAV, from 1.76 GPa (ZIF-4), 1.92 GPa (ZIF-62) and 4.20 GPa (TIF-4) respectively. Results for ZIF-4 are in approximate agreement with previous calculations and experimental results [7,29,35]. Across the ZIF-4, ZIF-62 and TIF-4 sequence, while nanoindentation values exceed calculated values by x2 [6], a similar trend of $E$ increasing with SAV was obtained.

Smaller differences are noted in the shear moduli of the three materials, viz. 1.11 GPa (ZIF-4), 1.33 GPa (ZIF-62) and 1.83 GPa (TIF-4) respectively, increasing with decreasing SAV. The data also reveal that the aggregate values of Poisson's ratio for all three ZIFs (Table 3) fall below the brittle-ductile transition of $v$~0.33 [10], which means that, outside the harmonic regime, these compliant 'soft' materials are liable to fracture, substantiating the hypothesis that mechanical amorphization of ZIF powders could be straightforwardly triggered beyond the yield point by shear-induced angular distortions. It is also important to note that the determined values of the aggregate Poisson's ratio are positive, whereas ZIF-4 and ZIF-62 exhibit a relatively strong auxetic response in single crystal form. If this auxeticity effect should prove useful, it's potential for developing growth methodology for manufacturing polycrystalline MOF films rests with accurately matching crystallographic orientations.

## 4  Conclusion

We have investigated the mechanical properties of three ZIFs materials with the equivalent topologies, by means of ab initio density functional theory (DFT). The elastic constants of three zeolitic imidazolate frameworks, namely, ZIF-4, ZIF-62 and TIF-4, have been calculated by the efficient stress-strain method at the optimized structures. While these three materials have the same topological structure and framework architecture, they incorporate different organic ligands. Our results have revealed the effect of mixing different organic ligands on their crucial mechanical characteristics, such as Young's modulus, shear modulus, Poisson's ratio or linear compressibility. All the structures exhibit low Young's moduli and shear moduli due to their compliant environment around cation nodes and molecular ligands. More interestingly, all of them display negative Poisson's ratio (NPR) and negative linear compressibility (NLC). In particular, this full tensorial analysis sheds light on the opportunities for better understanding and modulating their mechanical properties.

## Acknowledgements

This work is financially supported by the National Natural Science Foundation of China (Nos. 51461135004, 11604249), China Scholarship Council (CSC) under project number 201606955033, the Key Technology Innovation Project of Hubei Province (No. 2013AAA005), the Natural Science Foundation of Hubei Province (No. 2015CFB227), Scientific Leadership training Program of Hubei province (No. [2012]86), the Fundamental Research Funds for the Central Universities (Nos. 2015IVA051, 2017IVB020, 2017IIGX47, 2016-YB-009), and the research board of the State Key Laboratory of Silicate Materials for Architectures. We also thank Shanghai Supercomputer Center for providing computing resources. T. D Bennett and A. K. Cheetham are acknowledged for helpful discussions.

## References

[1] C. Rao, A. Cheetham, A. Thirumurugan,. Journal of Physics: Condensed Matter, 20 (2008) 083202.

[2] G. Férey. Dalton Transactions, (2009) 4400–4415.

[3] H. Furukawa, K. E. Cordova, M. O´ Keeffe, O. M. Yaghi, Science, 341 (2013) 1230444.

[4] J. E. Mondloch, M. J. Katz, W. C. Isley III, P. Ghosh, P. Liao, W. Bury, G. W. Wagner, M. G. Hall, J. B. DeCoste and G. W. Peterson, Nature materials, 14 (2015) 512–516.

[5] A. Phan, C. J. Doonan, F. J. Uribe-Romo, C. B. Knobler, M. O´keeffe and O. M. Yaghi, Acc. Chem. Res, 43 (2010) 58–67.

[6] J. C. Tan, T. D. Bennett and A. K. Cheetham,.Proceedings of the National Academy of Sciences, 107 (2010) 9938–9943.

[7] J.-C. Tan, B. Civalleri, A. Erba and E. Albanese, CrystEngComm, 17 (2015) 375–382.

[8] L. Bouëssel du Bourg, A. U. Ortiz, A. Boutin and F.-X. Coudert, APL materials, 2 (2014) 124110.

[9] A. U. Ortiz, A. Boutin, A. H. Fuchs and F.-X. Coudert, The Journal of Physical Chemistry Letters, 4 (2013) 1861–1865.

[10] G. N. Greaves, A. Greer, R. Lakes and T. Rouxel, Nature materials, 10 (2011) 823.

[11] Z. A. Lethbridge, R. I. Walton, A. S. Marmier, C. W. Smith and K. E. Evans, Acta Materialia, 58 (2010) 6444–6451.

[12] G. Greaves, F. Meneau, O. Majerus, D. Jones and J. Taylor, Science, 308 (2005) 1299–1302.

[13] G. N. Greaves, F. Meneau, F. Kargl, D. Ward, P. Holliman and F. Albergamo, Journal of Physics: Condensed Matter, 19 (2007) 415102.

[14] T. D. Bennett, Y. Yue, P. Li, A. Qiao, H. Tao, N. G. Greaves, Richards, T. Richards, G. I. Lampronti, S. A. Redfern and F. Blanc, Journal of the American Chemical Society, 138 (2016) 3484–3492.

[15] K. S. Park, Z. Ni, A. P. Côté J. Y. Choi, R. Huang, F. J. Uribe-Romo, H. K. Chae, M. O'Keeffe, O. M. Yaghi, Proceedings of the National Academy of Sciences, 103 (2006) 10186–10191.

[16] M. Gustafsson and X. Zou, Journal of Porous Materials, 20 (2013) 55–63.

[17]T. Wu, X. Bu, J. Zhang and P. Feng, Chemistry of Materials, 20 (2008) 7377–7382.

[18] T. D. Bennett, A. L. Goodwin, M. T. Dove, D. A. Keen, M. G. Tucker, E. R. Barney, A. K. Soper, E. G. Bithell, J.-C. Tan, A. K. Cheetham, Physical Review Letters, 104 (2010) 115503.

[19] T. D. Bennett, J.-C. Tan, Y. Yue, E. Baxter, C. Ducati, N. J. Terrill, H. H.-M. Yeung, Z. Zhou, W. Chen and S. Henke, Nature communications, 6 (2015).

[20] G. Kresse and J. Furthmüller, Phys. Rev. B, 54 (1996) 11169–11186.

[21] J. P. Perdew, K. Burke and M. Ernzerhof, Physical review letters, 77 (1996) 3865.

[22] Y. Le Page and P. Saxe, Physical Review B, 65 (2002) 104104.

[23] X. Wu, D. Vanderbilt and D. Hamann, Physical Review B, 72 (2005) 035105.

[24] H. Ledbetter and A. Migliori, Journal of Applied Physics, 100 (2006) 063516.

[25] W. C. Oliver and G. M. Pharr, Journal of materials research, 7 (1992) 1564–1583.

[26] J. F. Nye, Physical properties of crystals: their representation by tensors and matrices; Oxford university press, (1985).

[27] W. Voigt, BG Teubner, Leipzig, Germany, (1928).

[28] Z.-j. Wu, E.-j. Zhao, H.-p. Xiang, X.-f. Hao, X.-j. Liu and J. Meng, Physical Review B, 76 (2007) 054115.

[29] M. R. Ryder and J.-C. Tan, Dalton Transactions, 45 (2016) 4154–4161.

[30] J.-C. Tan, B. Civalleri, C.-C. Lin, L. Valenzano, R. Galvelis, P.-F. Chen, T. D. Bennett, C. Mellot-Draznieks, C. M. Zicovich-Wilson and A. K. Cheetham, Physical review letters, 108 (2012) 095502.

[31] M. R. Ryder, B. Civalleri, T. D. Bennett, S. Henke, S. Rudic, G. Cinque, F. Fernandez-Alonso and J.-C. Tan, Physical review letters, 113 (2014) 215502.

[32] A. U. Ortiz, A. Boutin, A. H. Fuchs and F.-X. Coudert, Physical review letters, 109 (2012) 195502.

[33] R. Hill, Proceedings of the Physical Society. Section A, 65 (1952) 349.

[34] S. I. Ranganathan and M. Ostoja-Starzewski, Physical Review Letters, 101 (2008) 055504.

[35] T. D. Bennett, P. Simoncic, S. A. Moggach, F. Gozzo, P. Macchi, D. A. Keen, J.-C. Tan and A. K. Cheetham, Chemical communications, 47 (2011) 7983–7985.

**Table 1** Structure properties of ZIF-4, ZIF-62 and TIF-4, Density = g/cm$^3$, T/V = density of metal atoms per nm$^3$; SAV = solvent accessible volume. The SAV were calculated using the "VOID" algorithm implemented in the PLATON package. Density (SOL) incorporates template density scaled by SAV. Densities (EXPT) are reported densities of solvated crystals [6]. ρ (DFT SOL) =ρ (DFT DESOL) + SAV/100* ρ (template).

| ZIFs | $a$(Å) | $b$(Å) | $c$(Å) | Density | SAV | T/V |
|---|---|---|---|---|---|---|
| ZIF-4[this work] | 15.350 | 15.811 | 18.519 | 1.503 | 34.3% | 3.56 |
| ZIF-4 Expt.[15] | 15.395 | 15.307 | 18.426 | 1.498 | | |
| ZIF-62[this work] | 14.914 | 17.398 | 19.078 | 1.498 | 31.0% | 3.23 |
| ZIF-62 Expt.[16] | 15.662 | 15.662 | 18.207 | 1.494 | | |
| TIF-4[this work] | 15.015 | 17.658 | 19.188 | 1.459 | 28.6% | 3.14 |
| TIF-4 Expt.[17] | 15.625 | 16.322 | 18.124 | 1.449 | | |

**Table 2** Single-crystal elastic stiffness coefficients of ZIF-4, ZIF-62 and TIF-4.

| $C_{ij}$ (GPa) | $C_{11}$ | $C_{22}$ | $C_{33}$ | $C_{44}$ | $C_{55}$ | $C_{66}$ | $C_{12}$ | $C_{13}$ | $C_{23}$ |
|---|---|---|---|---|---|---|---|---|---|
| ZIF-4 | 1.7232 | 3.7209 | 3.6638 | 0.74726 | 1.7456 | 2.4126 | 0.98411 | 1.5217 | 1.7698 |
| ZIF-62 | 1.1396 | 7.5038 | 6.162 | 1.4858 | 1.8465 | 1.7703 | 0.25711 | 0.41292 | 3.1753 |
| TIF-4 | 3.8971 | 11.385 | 6.1894 | 1.7544 | 2.1387 | 2.4466 | 2.9561 | 2.6158 | 5.464 |
| ZIF-4[7] | 4.266 | 3.492 | 5.015 | 1.029 | 1.927 | 2.453 | 1.221 | 1.916 | 1.526 |
| ZIF-4[32] | 3.073 | 3.361 | 2.953 | 0.771 | 0.903 | 1.532 | 0.574 | 0.603 | 0.770 |

**Table 3** Mechanical properties of ZIF-4, ZIF-62 and TIF-4

| Elastic property | | ZIF-4 | ZIF-62 | TIF-4 |
|---|---|---|---|---|
| Young's modulus | $E_{max}$ | 3.439 | 4.8438 | 6.4114 |
| E(GPa) | $E_{min}$ | 1.0696 | 0.76313 | 2.7278 |
| | $A_E$ | 3.215 | 6.347 | 2.35 |
| Shear modulus, | $G_{max}$ | 2.4126 | 1.8465 | 2.4466 |
| G(GPa) | $G_{min}$ | 0.4722 | 0.4070 | 1.0703 |
| | $A_G$ | 5.109 | 4.537 | 2.286 |
| Linear compressibility | $\beta_{max}$ | 504.63 | 1101.2 | 207.43 |
| (β TPa$^{-1}$) | $\beta_{min}$ | -2.5994 | -204.69 | -2.5881 |
| Poisson's ratio, | $\nu_{max}$ | 0.72019 | 1.0802 | 0.78487 |
| (ν) | $\nu_{min}$ | -0.49726 | -0.79459 | -0.049 |

**Table 4** isotropic aggregate elastic properties based on the Voigt-Reuss-Hill (VRH) averages, corresponding to a texture-free polycrystalline material. $A^U$ represents the universal elastic anisotropy.

| ZIFs | $K_{VRH}$(GPa) | $E_{VRH}$(GPa) | $G_{VRH}$(GPa) | $\nu_{VRH}$(GPa) | $A^U$ |
|---|---|---|---|---|---|
| ZIF-4 | 1.76 | 2.75 | 1.11 | 0.24 | 2.36 |
| ZIF-62 | 1.92 | 3.24 | 1.33 | 0.22 | 5.00 |
| TIF-4 | 4.20 | 4.79 | 1.83 | 0.31 | 1.15 |

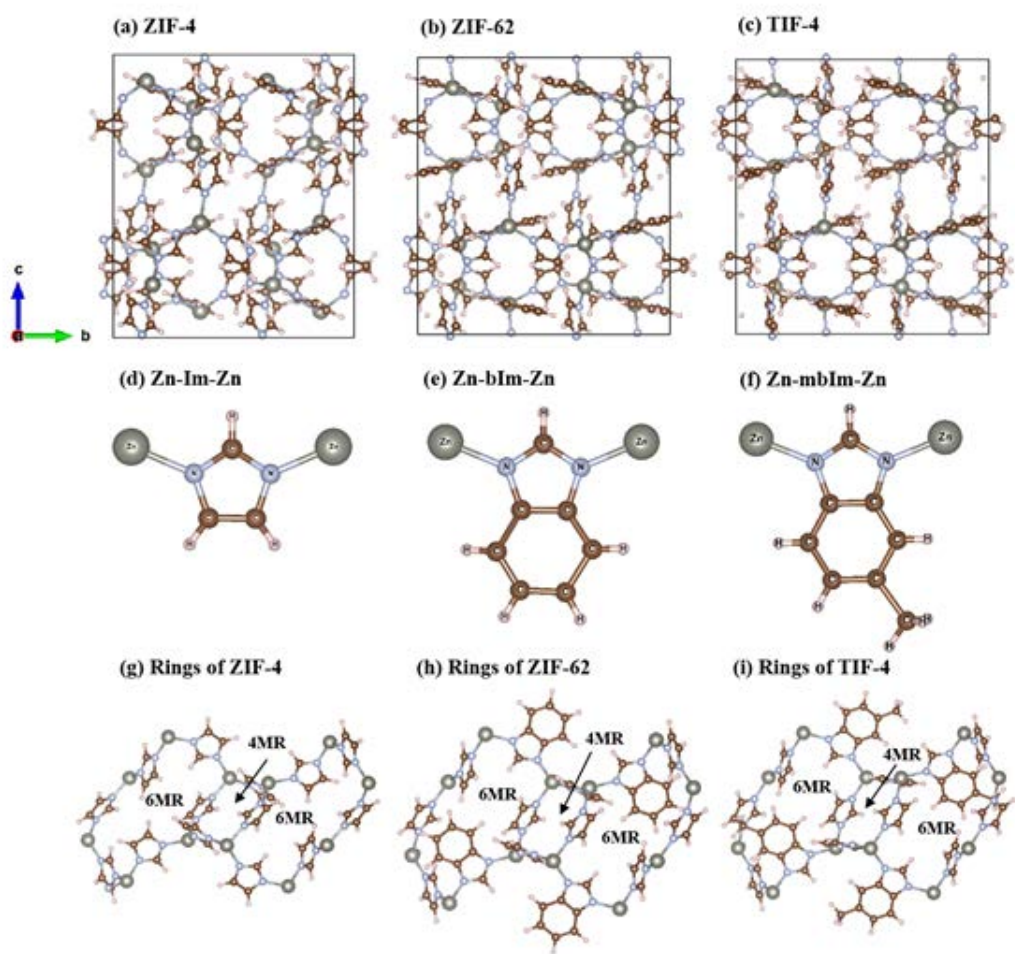

**Fig. 1** Crystal structures of (a) ZIF-4, (b) ZIF-62, and (c) TIF-4. The different zinc-organic ligands-zinc structures of (d) Zn-Im-Zn, (e) Zn-bIm-Zn, and (f) Zn-mbIm-Zn. 4 and 6 member ring structure of (g) ZIF-4, (h)ZIF-62, (i)TIF-4, Color labels: zinc: grey, carbon: brown, nitrogen: light blue, hydrogen: pink

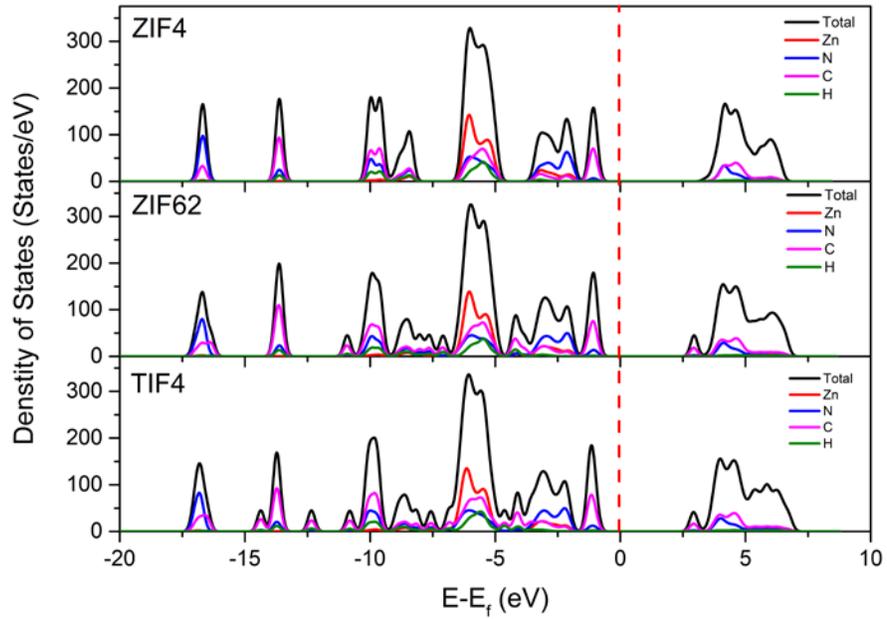

**Fig. 2** Calculated total and partial density of states of ZIF-4, ZIF-62 and TIF-4

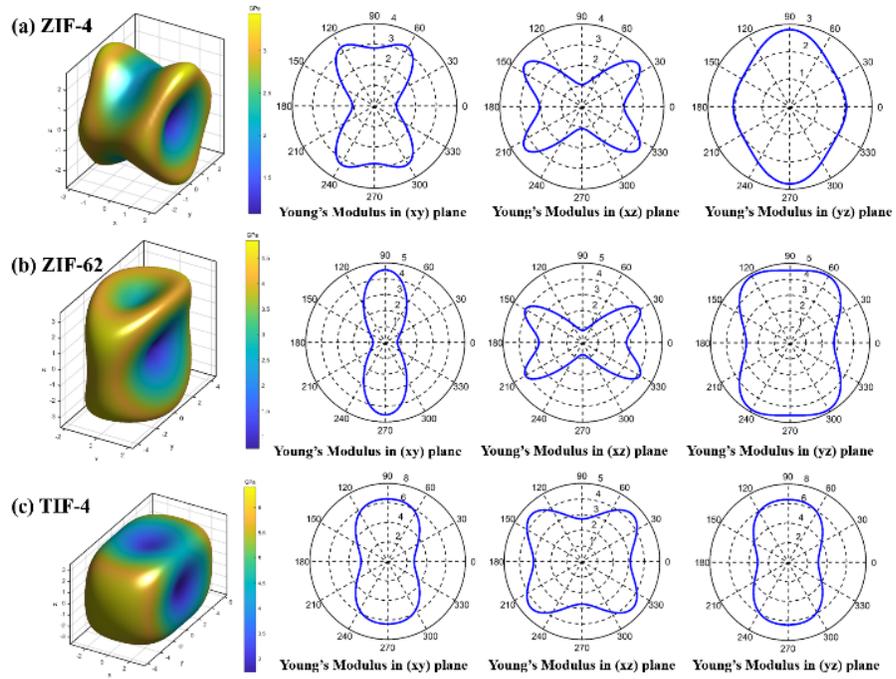

**Fig. 3** 3D and 2D representations of Young's modulus for (a) ZIF4, (b) ZIF62 and (c) TIF4.

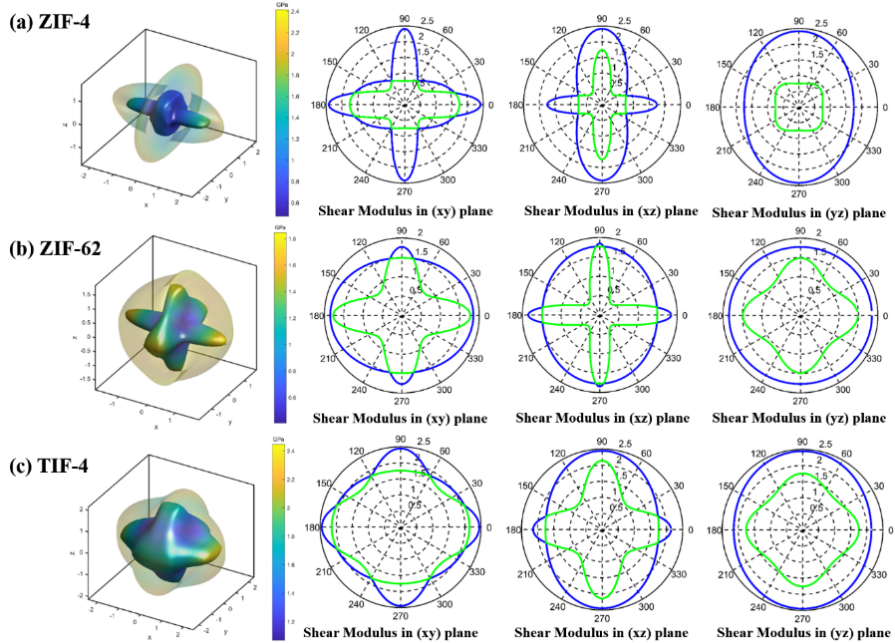

**Fig. 4** 3-D and 2-D representation surface of Shear modulus G for (a) ZIF-4, (b) ZIF62 and (c) TIF4. The maximum and minimum values are represented as blue and green surfaces, respectively.

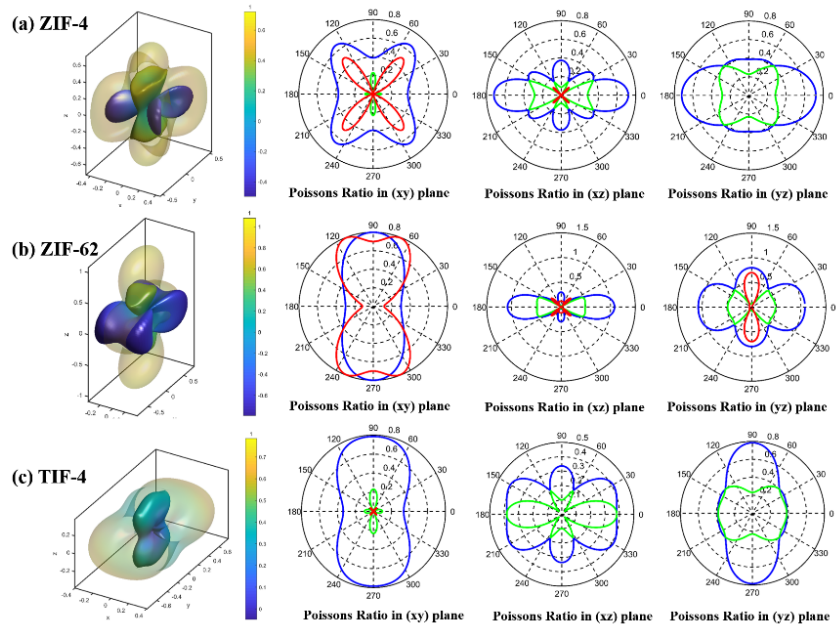

**Fig. 5** 3D and 2D representation surfaces of Poisson's ratio for (a)ZIF-4, (b) ZIF-62 and (c) TIF-4. The blue line signifies the maximum υ. The minimum υ presented in green denotes a positive minimum, while red is used to represent a negative minimum or an auxetic response.

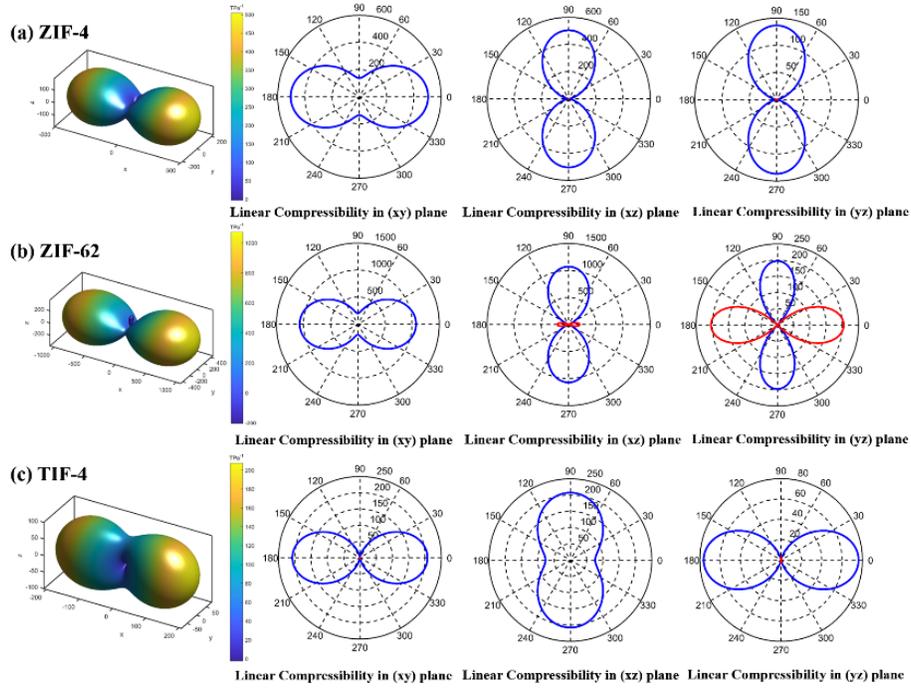

**Fig. 6** 3D and 2D representation surfaces of linear compressibility representation for (a)ZIF-4, (b) ZIF-62 and (c) TIF-4. The green and red lines label positive and negative compressibility, respectively.